\documentclass{article}

\PassOptionsToPackage{numbers, compress}{natbib}
\usepackage[preprint]{neurips_2023}

\usepackage[utf8]{inputenc} 
\usepackage[T1]{fontenc}    
\usepackage{hyperref}       
\usepackage{url}            
\usepackage{booktabs}       
\usepackage{amsfonts}       
\usepackage{nicefrac}       
\usepackage{microtype}      
\usepackage{xcolor}         

\usepackage{amsmath,amssymb,amsfonts}
\usepackage{algorithmic}
\usepackage{graphicx}
\usepackage{textcomp}
\usepackage{graphicx}
\usepackage{caption}
\usepackage{subcaption}
\usepackage{bbm}
\usepackage{bm}
\usepackage{color}
\usepackage{hyperref}
\usepackage{multirow}
\usepackage{enumitem}
\usepackage{algorithm} 
\usepackage{algorithmic}
\usepackage{enumerate}
\usepackage{amsmath}
\usepackage{booktabs}
\usepackage{multirow}
\usepackage{stmaryrd}
\usepackage{color}
\usepackage{balance}
\usepackage{adjustbox}
\usepackage{mathtools}

\title{Towards Efficient and Effective Unlearning of Large Language Models for Recommendation}

\author{%
    Hangyu Wang \thanks{Equal contribution.} \\
    Shanghai Jiao Tong University \\
    Shanghai, China \\
    hangyuwang@sjtu.edu.cn \\
    \And 
    Jianghao Lin \footnotemark[1] \\
    Shanghai Jiao Tong University \\
    Shanghai, China \\
    chiangel@sjtu.edu.cn \\
    \And 
    Bo Chen \\
    Huawei Noah’s Ark Lab \\
    Shenzhen, China \\
    chenbo116@huawei.com \\
    \And 
    Yang Yang \\
    Huawei Noah’s Ark Lab \\
    Shenzhen, China \\
    yangyang590@huawei.com \\
    \And 
    Ruiming Tang \\
    Huawei Noah’s Ark Lab \\
    Shenzhen, China \\
    tangruiming@huawei.com \\
    \And 
    Weinan Zhang \\
    Shanghai Jiao Tong University \\
    Shenzhen, China \\
    wnzhang@sjtu.edu.cn \\
    \And 
    Yong Yu \thanks{Corresponding author.} \\
    Shanghai Jiao Tong University \\
    Shanghai, China \\
    yyu@sjtu.edu.cn \\
}

\begin{document}

\maketitle

\begin{abstract}
The significant advancements in large language models (LLMs) give rise to a promising research direction, i.e., leveraging LLMs as recommenders (LLMRec). The efficacy of LLMRec arises from the open-world knowledge and reasoning capabilities inherent in LLMs. LLMRec acquires the recommendation capabilities through instruction tuning based on user interaction data. However, in order to protect user privacy and optimize utility, it is also crucial for LLMRec to intentionally forget specific user data, which is generally referred to as recommendation unlearning. In the era of LLMs, recommendation unlearning poses new challenges for LLMRec in terms of \textit{inefficiency} and \textit{ineffectiveness}. Existing unlearning methods require updating billions of parameters in LLMRec, which is costly and time-consuming. Besides, they always impact the model utility during the unlearning process. To this end, we propose \textbf{E2URec}, the first \underline{E}fficient and \underline{E}ffective \underline{U}nlearning method for LLM\underline{Rec}. Our proposed E2URec enhances the unlearning efficiency by updating only a few additional LoRA parameters, and improves the unlearning effectiveness by employing a teacher-student framework, where we maintain multiple teacher networks to guide the unlearning process. Extensive experiments show that E2URec outperforms state-of-the-art baselines on two real-world datasets. Specifically, E2URec can efficiently forget specific data without affecting recommendation performance. The source code is at \url{https://github.com/justarter/E2URec}.
\end{abstract}

\section{Introduction}
\begin{figure}
    \centering
    \includegraphics[width=0.7\linewidth]{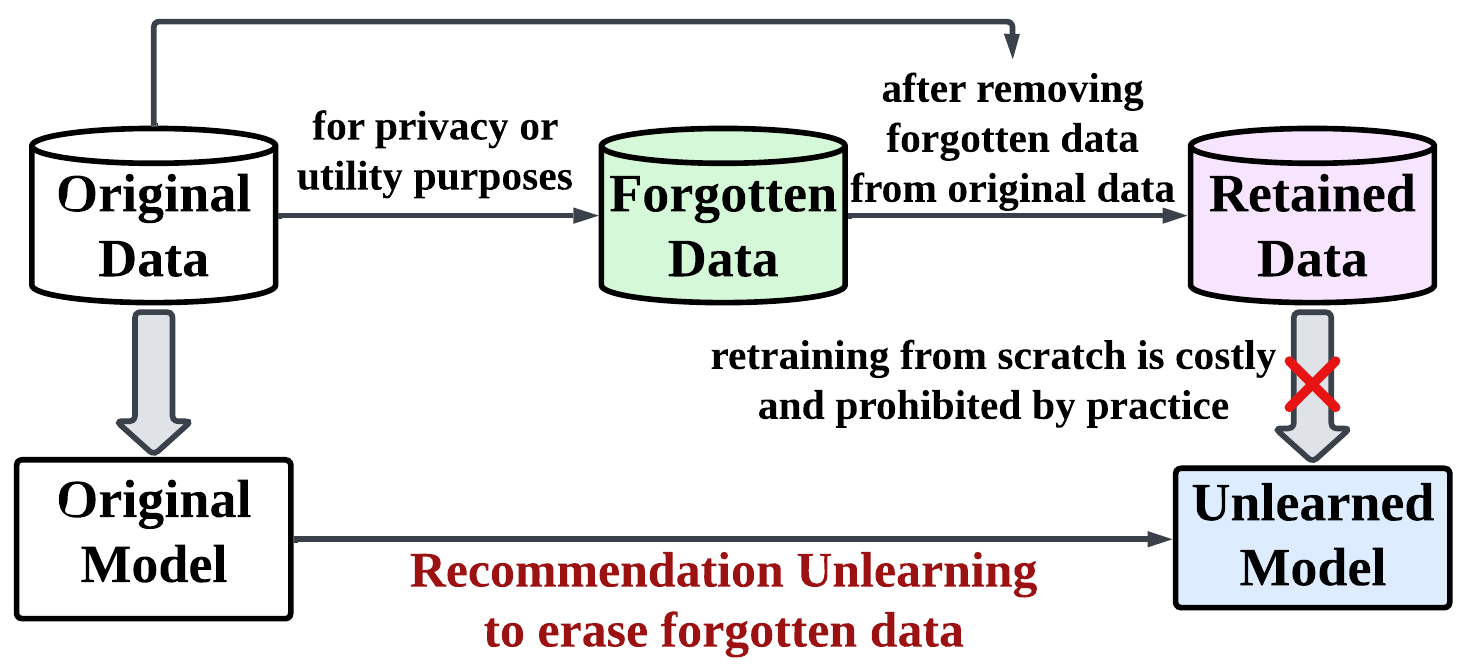}
    \caption{The illustration of recommendation unlearning. We aim to forget the recommendation knowledge about the chosen forgotten data due to privacy or utility purposes, and meanwhile keep the recommendation performance of the unlearned model for the rest data.}
    \label{fig:process}
\end{figure}

Large Language Models (LLMs) possess massive parameters and are trained on vast datasets, demonstrating exceptional proficiency in various tasks~\cite{t5}. 
The remarkable advancements in LLMs also inspire the exploration of leveraging LLMs as recommenders (LLMRec), whose effectiveness stems from extensive open-world knowledge and reasoning ability in LLMs~\cite{fan2023recommender,wu2023survey}.
LLMRec obtains the recommendation ability through instruction tuning on the user interaction data. 
But in many cases, it is also crucial for LLMRec to forget specific user data, which is referred to as \textbf{recommendation unlearning}~\cite{sisa}. As shown in Figure~\ref{fig:process}, recommendation unlearning aims to forget the recommender's knowledge about certain data (named as forgotten data) while ensuring that the original recommendation performance remains unaffected.

The necessity of recommendation unlearning mainly arises from two aspects: 1) Privacy. According to privacy legislation, such as the General Data Protection Regulation (GDPR)~\cite{gdpr} and the California Consumer Privacy Act (CCPA)~\cite{ccpa}, recommenders are obligated to erase the sensitive data upon user requests in order to protect user privacy.
2) Utility. Noisy data or polluted data can severely degrade recommendation performance. Once these data are identified, recommenders need to forget them to regain the utility~\cite{receraser}.

Although attempts have been made in recent studies~\cite{receraser,SCIF,ye2023sequence} to apply machine unlearning to recommender systems due to concerns about data privacy, they have predominantly focused on conventional small-parameter recommenders. These approaches require model retraining or Hessian matrix computation, which is expensive and time-consuming given the billions of parameters in LLMs. While some studies~\cite{yao2023large,chen2023unlearn} have explored unlearning in large language models, they are primarily designed in the Natural Language Processing (NLP) domain. Furthermore, these methods attempt to employ a gradient ascent-based approach to unlearn knowledge. While these approaches may yield effectiveness, they often impact the model utility on normal data~\cite{yao2023large}. As a result, how to efficiently and effectively erase the influence of specific data is not trivial in LLMRec.





To this end, we propose E2URec, \underline{E}fficient and \underline{E}ffective \underline{U}nlearning for LLM\underline{Rec}. For \textit{efficiency}, our proposal involves integrating a lightweight unlearning module into the original LLM, drawing inspiration from low-rank adaptation (LoRA)~\cite{hu2021lora}. Consequently, during the unlearning process, only the parameters of the lightweight LoRA module will be updated, while the parameters of the LLM remain frozen. For \textit{effectiveness}, we employ the teacher-student mode. Specifically, we introduce a forgetting teacher model that has never seen the forgotten data. During the unlearning process, the student (i.e., unlearned model) will approach the forgetting teacher, thus helping the student to forget the specific samples. We also devise a remembering teacher to guide the student to retain the original recommendation performance.

In conclusion, we summarize our main contributions as follows: 
\begin{itemize}
    \item To the best of our knowledge, we are the first to study the unlearning problem for LLMRec. Our proposed E2URec outperforms existing approaches in terms of both efficiency and effectiveness.
    \item For efficiency, we propose to add a lightweight low-rank adaptation (LoRA) module to the original LLM. Only the LoRA parameters will be updated in the unlearning, while the parameters of the LLM are frozen.
    \item For effectiveness, we design a novel forgetting teacher and a remembering teacher to guide the unlearned model, so that the unlearned model can forget data and maintain the recommendation performance respectively.
    \item Extensive experiments on two real-world public datasets demonstrate the superiority of our proposed E2URec in recommendation unlearning for LLMRec compared with existing baselines, in terms of both efficiency and effectiveness.
\end{itemize}

\section{Related Work}
\subsection{Machine Unlearning and Downstream Applications} 

We would like to first introduce the concept and categorization of general machine unlearning methods (i.e., exact unlearning and approximate unlearning), and then elaborate on their special downstream applications for both recommendation unlearning and large language model unlearning.

\textbf{Machine Unlearning} involves eliminating the influence of particular training data from a trained model~\cite{cao2015towards,nguyen2022survey}. One naive method to accomplish this is by retraining the model from scratch using a revised dataset that excludes the forgotten data. However, this method can be impractical due to its computational demands. Based on the degree of unlearning completeness, existing unlearning techniques can be categorized into exact unlearning and approximate unlearning.

\textit{Exact unlearning} requires eliminating all information relevant to the forgotten data so that the unlearned model performs identically to a completely retrained model. For example, SISA~\cite{sisa}, as a prominent approach, only requires retraining the sub-model trained on the relevant data shard for each unlearning request. GraphEraser~\cite{GraphEraser} extends SISA for unlearning in graph neural networks, while ARCANE~\cite{yan2022arcane} partitions the dataset by class and employs one-class anomaly detection training on each subgroup. However, these methods require model retraining, which is computationally expensive and time-consuming.

\textit{Approximate unlearning} only ensures that the performance of the unlearned model closely aligns with that of a retrained model~\cite{guo2019certified,sekhari2021remember}. These methods estimate the influence of the forgotten samples and mitigate it by inverse gradient-based update techniques~\cite{ng,randomlabel}, allowing efficient forgetting without retraining. This trade-off enhances efficiency and model utility but compromises completeness, leading to statistical forgetting~\cite{izzo2021approximate}. Certain studies use the influence function to quantify the influence of the forgotten data~\cite{koh2017understanding}, which entails computing the Hessian matrix and incurs additional computational overheads~\cite{mehta2022deep}. Recent research also highlights the computational intractability of the influence of individual training data on deep models~\cite{graves2021amnesiac}.

While machine unlearning techniques are widespread and well studied, many research works start to explore their applications for various downstream areas, e.g., recommendation unlearning and large language model unlearning.

\textbf{Recommendation Unlearning} has garnered increasing research interest due to both privacy and utility purposes~\cite{sachdeva2024machine,schelter2023forget,li2023making,ganhor2022unlearning}. Unlearning in recommender systems not only aids in safeguarding user privacy but also enhances recommendation models by mitigating the impact of noisy data~\cite{receraser}. RecEraser~\cite{receraser}, along with LASER~\cite{laser} and UltraRE~\cite{ultrare}, proposes novel data partition algorithms that divide the training set into balanced shards to preserve the collaborative information. AltEraser~\cite{alteraser} enhances the efficiency by dividing the optimization problem of unlearning into many small tractable sub-problems. SCIF~\cite{SCIF} and IFRU~\cite{IFRU} extend influence function by selectively updating user embedding and employing importance-based pruning algorithms, respectively. Additionally, IMCorrect~\cite{imcorrect} and Unlearn-ALS~\cite{Unlearn-ALS} improve exact unlearning by correcting the mapping matrix and using Alternating Least Squares optimization, respectively. Lastly, unlearning methods have also emerged specifically for sequential recommendation~\cite{ye2023sequence}, session-based recommendation~\cite{xin2023effectiveness} and federated recommendation~\cite{yuan2023federated}. However, none of the existing methods are tailored for LLM-based recommendation. Moreover, current approaches primarily prioritize forgetting effectiveness, neglecting the significance of preserving the original recommendation performance.

\textbf{Large Language Model Unlearning}. While there's a growing body of research on machine unlearning, LLM unlearning is still a largely under-explored topic~\cite{si2023knowledge,liu2024rethinking}. The extensive parameters and training data of LLMs render existing unlearning methods ill-suited and inefficient, since they are typically tailored for classification tasks and demand either model retraining or Hessian matrix computations. \citet{yao2023large} formulate the settings and goals in LLM unlearning, advocating the removal of harmful content using a Gradient Ascent (GA) approach. \citet{chen2023unlearn} introduce EUL, an efficient unlearning framework employing unlearning layers. Besides, \citet{eldan2023s} propose a novel network to unlearn copyrighted knowledge embedded within LLMs. However, they all use gradient ascent methods to unlearn knowledge, which will deteriorate the model performance on normal data~\cite{liu2024towards}. Lastly, two benchmark~\cite{maini2024tofu,yao2024machine} are established to evaluate the efficacy of various unlearning methods for LLMs. To the best of our knowledge, we are the first to study the LLM unlearning in the recommendation scenario. To overcome the shortcomings of gradient ascent methods, we adopt the teacher-student mode to maintain the original recommendation performance.

\subsection{Large Language Models as Recommenders}

With the remarkable development of large language models~\cite{zhao2023survey}, researchers start to investigate their potential to directly perform the recommendation tasks, i.e., large language models as recommenders (LLMRec). As suggested in previous work~\cite{lin2023can,xu2024prompting}, according to different tasks to be solved by LLMs, LLMRec can be generally classified in three categories: (1) item scoring task, (2) item generation task, and (3) hybrid task.

In the \textit{item scoring} task~\cite{zhang2023collm,li2023e4srec,wang2023flip,lin2023clickprompt,bao2023tallrec,lin2023rella}, LLMs are required to perform pointwise preference estimation for each user-item pair, and the final ranked item list is generated by sorting the estimated preference scores. 
In the \textit{item generation} task~\cite{liao2023llara,hou2023large,zheng2023adapting,zhang2023agentcf}, LLMs serve as generative functions to directly produce the final ranked list. Generally speaking, such a generative task places a significant demand on the powerful reasoning and instruction-following capabilities of LLMs to infer the user preference for item generation.
As for the \textit{hybrid} task~\cite{zhang2023recommendation,geng2022recommendation,luo2023recranker}, LLMs would accomplish the recommendation in a multi-task manner which combines both the item scoring and generation tasks. The basis to support such hybrid functionality is that LLMs are inherent multi-task learners~\cite{brown2020language,ouyang2022training} through a unified language interface.

In this paper, we mainly focus on the first paradigm of LLMRec (i.e., item scoring task), which is more similar and aligned with traditional recommendation tasks. However, our proposed E2URec is generalized, and is also capable of item generation and hybrid tasks, which we leave as future works.

\section{Preliminary}
\subsection{LLMs as Recommenders}
Conventional recommendation aims to precisely estimate the user preference for each candidate item, and finally arrange a ranked list presented to users~\cite{lin2023map,xi2023bird, huang2022neural}.
LLMRec (LLMs as Recommenders) aims to utilize LLM to predict whether a new item will be clicked by a user. Here, click denotes the positive interaction and can be replaced by behaviors such as purchase or like. We denote the recommendation dataset as $\mathcal{D}=\{(x_i,y_i)\}_{i=1}^N$ with $N$ samples, where $x_i$ is the features of $i$-th sample, and $y_i$ is the label. Features $x_i$ is transformed into textual sentence $x_i^{text}$ via hard prompt templates. 
For example, $x_i^{text}$ may be: 
\begin{equation}
  \parbox{\dimexpr\linewidth-4em}{%
    \strut
    $\bm{x_i^{text}}=\;\,\,$``\textit{The user watched the following movies in order: Pump UP the Volume, Anta, Devil's Own, Crying Game. Please deduce if he will like the movie Titanic.}''
    \strut
  }
  \noindent
\end{equation}
Similarly, the binary label $y_i \in \{1,0\}$ (click or not) is converted into corresponding binary answer words $y_i^{text} \in \{\text{``Yes''},\text{``No''}\}$. 

In order to tailor LLM to recommendation scenarios, the causal language modeling objective~\cite{bao2023tallrec} is used to optimize LLM on dataset $\mathcal{D}$, by minimizing the negative log-likelihood of generating $y_i^{text}$ conditioned on input $x_i^{text}$: 
\begin{equation}
    L_{pred}(\mathcal{D}) = - \sum_{(x_i^{text}, y_i^{text}) \in \mathcal{D}} \sum_{t=1}^{|y|} \log \left(P\left(y^{text}_{i,t} \big\vert x^{text}_i,y^{text}_{i,<t}\right)\right)
    \label{eq:recommendationloss}
\end{equation}
where $y_{i,t}^{text}$ is the $t$-th token of $y_i^{text}$, and $y_{i,<t}^{text}$ is the tokens before $y_{i,t}^{text}$.

\subsection{Recommendation Unlearning for LLMRec}
Formally, suppose that on the recommendation dataset $\mathcal{D}$, we have trained a LLMRec model that is called the \textbf{original model}.
Subsequently, a request for deletion is received, whereby the data to be removed is denoted as \textbf{forgotten data} $\mathcal{D}_f$, and the remaining data is \textbf{retained data} $\mathcal{D}_r=\mathcal{D}- \mathcal{D}_f$. 
Then, the goal of unlearning is to learn a \textbf{unlearned model} $\mathcal{M}_u$ that forgets the information about $\mathcal{D}_f$ without hurting the recommendation performance. The unlearned model is initialized with the parameters of the original model.

\section{Methodologies}

\subsection{Parameter-Efficient Unlearning}

\begin{figure}
    \centering
    \includegraphics[width=0.7\linewidth]{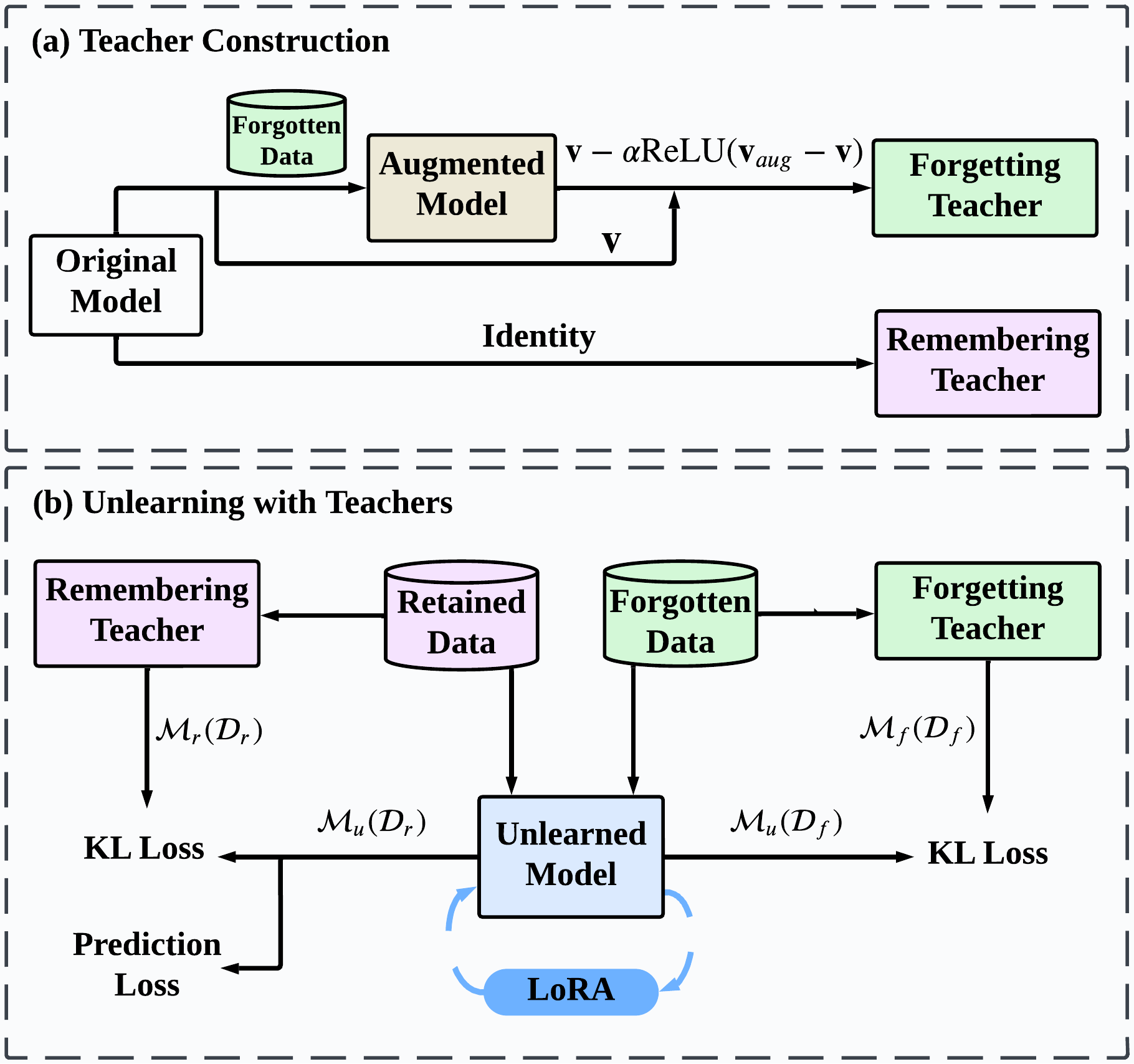}
    \caption{The overall framework of our proposed E2URec, which consists of two major stages: (1) Teacher Construction, and (2) Unlearning with Teachers. }
    \label{fig:framework}
\end{figure}

Considering the billions of parameters of LLM, updating all its parameters for forgetting is resource-intensive. Inspired by recent advances in parameter-efficient finetuning~\cite{hu2021lora}, we propose to insert the lightweight LoRA modules into LLM, as shown in Figure~\ref{fig:framework}(b). The LoRA modules add pairs of rank decomposition weight matrices to the original parameters of the LLM while just introducing a few parameters.
In this way, we model the unlearned model as $\mathcal{M}_u(\cdot; \phi, \theta)\coloneqq \mathcal{M}_u(\cdot; \theta)$, where $\phi$ is the parameters of LLM and $\theta$ is the LoRA parameters. During the unlearning process, we only need to update $\theta$, while $\phi$ remains frozen. This greatly reduces the computing resource and time.

\subsection{Unlearning with Forgetting/Remembering Teachers}

We aim to achieve unlearning by using two teachers, as depicted in Figure~\ref{fig:framework}(b). To remove knowledge, we update the unlearned model $\mathcal{M}_u$ to produce distributions similar to the forgetting teacher $\mathcal{M}_f$ on the forgotten data $\mathcal{D}_f$. Simultaneously, to preserve recommendation performance, we update the unlearned model $\mathcal{M}_u$ to produce distributions similar to the remembering teacher $\mathcal{M}_r$ on the retained data $\mathcal{D}_r$. The whole process can be formulated as:
\begin{equation}
\begin{aligned}
&\min_{\theta}\ \text{KL}\Big( \mathcal{M}_f \Big(\mathcal{D}_f \Big) \ \Big\Vert \ \mathcal{M}_u \Big(\mathcal{D}_f;\theta \Big) \Big) \\
&\min_{\theta}\ \text{KL} \Big( \mathcal{M}_r \Big(\mathcal{D}_r \Big)\ \Big\Vert \  \mathcal{M}_u \Big(\mathcal{D}_r;\theta \Big) \Big) 
\end{aligned}
\end{equation}
where $\text{KL}(\cdot)$ is the KL divergence between the output probability distributions of the teacher and unlearned model. 

The forgetting teacher should have never seen the forgotten data. The retrained model, which refers to the model trained from scratch without observing $\mathcal{D}_f$, seems to be a suitable forgetting teacher. However, it is inefficient and not viable in practice. Here, we use an approximate method to design the forgetting teacher. 

As shown in Figure~\ref{fig:framework}(a), we first finetune an augmented model on the forgotten data $D_f$. The augmented model with additional training on $D_f$ will output logits that are more relevant to $D_f$. Therefore, the difference between the logits of the augmented and the original model represents the information related to the forgotten data $D_f$.
We denote the logits of the augmented model and the original model as $\mathbf{v}_{aug}$ and $\mathbf{v}$ respectively, so the difference is
$\mathbf{v}_{aug} - \mathbf{v}$.
Subtracting this difference from the logits of original model can obtain the logits $\mathbf{v}_{f}$, which exclude the $D_f$ information. The formula of logits $\mathbf{v}_{f}$ is as follows:
\begin{equation}
    \mathbf{v}_{f} = \mathbf{v} - \alpha \text{ReLU} (\mathbf{v}_{aug} - \mathbf{v})
\end{equation}
where $\alpha$ is a positive hyper-parameter.
Then, the output probability distribution of the forgetting teacher is the normalized $\mathbf{v}_{f}$, defined as $\text{Softmax}(\mathbf{v}_{f})$.

So far, we have acquired the forgetting teacher's outputs. Then the forgetting loss can be formulated as:
\begin{equation}
    L_{FGT} = \sum_{x_f \in \mathcal{D}_f}\ \text{KL}\left(\mathcal{M}_f\left(x_f\right) \Big\Vert \mathcal{M}_u \left(x_f;\theta \right)\right)
\end{equation}

Simply forgetting will hurt the model's recommendation performance. To retain the original recommendation ability, we encourage the unlearned model to ``stay close'' to the remembering teacher on retain data. We choose the original model as the remembering teacher $\mathcal{M}_r$, because it has the best recommendation performance. Besides, to further strengthen the knowledge related to the recommendation task, we also add the prediction loss from Equation~\ref{eq:recommendationloss}. Formally, the remembering loss is:
\begin{equation}
    L_{REM} = L_{pred} (\mathcal{D}_r;\theta) + \sum_{x_r \in \mathcal{D}_r}\ \text{KL}\left(\mathcal{M}_r\left(x_r\right) \Big\Vert \mathcal{M}_u\left(x_r;\theta \right)\right) 
\end{equation}

In conclusion, the purpose of constructing forgetting teacher is to efficiently simulate the retrained model after removing forgotten data. Therefore, learning from the forgetting teacher is to make the unlearned model learn how to forget data. On the other hand, the remembering teacher is actually the original model, so learning from the remembering teacher is to make the unlearned model maintain the recommendation performance. Finally, the loss of E2URec is the weighted sum of forgetting loss and remembering loss controlled by the hyper-parameter $\beta$:
\begin{equation}
    L = \beta L_{FGT} + (1-\beta) L_{REM}
\end{equation}

\section{Experiments}

\subsection{Experiment Setups}
We conduct experiments on two public recommendation datasets: MovieLen-1M (ML-1M) and GoodReads (GD). 
Due to the large scale of datasets, we downsample each datasets to 100,000 samples. 
Both datasets are split into training, validation and testing sets with a ratio of 6:2:2 according to the global timestamp. In the experiment, 20\% randomly chosen users would request to remove their training data. We use T5-base~\cite{t5} as the LLM backbone. We set $\alpha=2$ and $\beta=0.6$. \textit{Our method only needs to update 0.7\% of the total parameters}.

We compare our E2URec with the state-of-the-art methods:
\textbf{Original}: the original model without unlearning. \textbf{Retrain}: the model retrained from scratch without the forgotten data. We include it as a gold standard. \textbf{SISA}~\cite{sisa}: Sharded, Isolated, Sliced and Aggregated training. \textbf{RecEraser}~\cite{receraser}: improves SISA by collaborative sharding and aggregation. 
\textbf{NegKL}:
uses KL loss to finetune the original model both on the retained and forgotten data, negating the KL loss for the latter.
\textbf{NegGrad}~\cite{ng}: 
uses prediction loss to finetune the original model both on the retained and forgotten data, negating the gradient for the latter.
\textbf{Bad-T}~\cite{randomlabel}: 
use prediction loss to finetune the original model both on the retained and forgotten data, randomly assigning arbitrary labels for the latter.

\begin{table}
    \centering
    \begin{footnotesize}
    \caption{All metrics comparison results (in \%) on MovieLens-1M dataset. The best results (except for original and retrain) are in bold.}
    \label{tab:mainres_ml}
    \resizebox{\linewidth}{!}{
    \renewcommand\arraystretch{1}
    \begin{tabular}{cccccccc}
    \toprule
         \multirow{2}{*}{\textbf{Metrics}} & \multicolumn{5}{c}{\textbf{Effectiveness}} & \multicolumn{2}{c}{\textbf{Efficiency}} \\ \cmidrule(r){2-6} \cmidrule(r){7-8}
         & \textbf{AUC} $\uparrow$ & \textbf{ACC} $\uparrow$ & \textbf{LL} $\downarrow$ & \textbf{JSD} $\downarrow$ & \textbf{L2-norm} $\downarrow$ & \textbf{Time(s)} $\downarrow$ & \textbf{\#Params} $\downarrow$ \\ \midrule
       Original  & 77.44 & 70.60 & 56.78 & - & - & 9048 & $2.2\times 10^8$ \\
       Retrain & 76.85 & 69.98 & 57.35 & - & - & 5279 & $2.2\times 10^8$ \\ \midrule
       SISA & 75.35 & 68.52 & 58.89 & 2.05 & 9.82 & 3042 & $8.9\times 10^8$ \\
       RecEraser & 75.59 & 68.84 & 58.86 & 2.03 & 9.64 & 4009 & $8.9\times 10^8$\\
       NegKL & 75.65 & 69.19 & 59.34 & 3.67 & 12.47 & 1805 & $2.2\times 10^8$\\
       NegGrad & 75.97 & 69.31 & 59.20 & 4.64 & 14.13 & 1940 & $2.2\times 10^8$\\
       Bad-T & 75.61 & 69.41 & 58.83 & 4.35 & 14.95 & 1684 & $2.2\times 10^8$\\
       E2URec & \textbf{76.34} & \textbf{69.76} & \textbf{57.75} & \textbf{1.91} & \textbf{9.51} & \textbf{941} & \boldsymbol{$1.7\times 10^6$}\\
       \bottomrule
    \end{tabular}
    }
    \end{footnotesize}
\end{table}

\begin{table}
    \centering
    \begin{footnotesize}
    \caption{All metrics comparison results (in \%) on GoodReads dataset. The best results (except for original and retrain) are in bold.}
    \label{tab:mainres_gd}
    \resizebox{\linewidth}{!}{
    \renewcommand\arraystretch{1}
    \begin{tabular}{cccccccc}
    \toprule
    \multirow{2}{*}{\textbf{Metrics}} & \multicolumn{5}{c}{\textbf{Effectiveness}} & \multicolumn{2}{c}{\textbf{Efficiency}} \\ \cmidrule(r){2-6} \cmidrule(r){7-8}
         & \textbf{AUC} $\uparrow$ & \textbf{ACC} $\uparrow$ & \textbf{LL} $\downarrow$ & \textbf{JSD} $\downarrow$ & \textbf{L2-norm} $\downarrow$ & \textbf{Time(s)} $\downarrow$ & \textbf{\#Params} $\downarrow$ \\ \midrule
       Original  & 73.52 & 70.67 & 55.46 &  - & - & 9152 & $2.2\times 10^8$ \\
       Retrain & 73.39 & 70.53 & 55.56 & - & - & 5448 & $2.2\times 10^8$ \\ \midrule
       SISA & 72.19 & 70.07 &56.76 &  2.04 & 8.85 & 3008 & $8.9\times 10^8$ \\
       RecEraser & 72.29 & 69.93 &56.57 &  1.66 & 7.71 & 3208 & $8.9\times 10^8$ \\
       NegKL & 72.88 & 70.15 & 56.38 & 2.02 & 10.01 & 1866 & $2.2\times 10^8$ \\
       NegGrad & 72.85 &  70.26 & 57.21 &2.56 & 10.44 & 1608 & $2.2\times 10^8$ \\
       Bad-T & 72.75 & 70.13 &61.43 &  8.02 & 19.09 & 1753 & $2.2\times 10^8$ \\
       E2URec & \textbf{73.41} & \textbf{70.42} & \textbf{55.48} &  \textbf{0.90} & \textbf{6.54} & \textbf{800} & \boldsymbol{$1.7\times 10^6$} \\
       \bottomrule
    \end{tabular}
    }
    \end{footnotesize}
\end{table}

We use the following metrics for analysis: 1) \textit{AUC, ACC and LogLoss (LL) on test set}: measure the recommendation performance of the unlearned model. 2) \textit{JS-Divergence (JSD) and L2-norm on the forgotten data}: JSD and L2-norm between the outputs of the unlearned and retrained model measure the effectiveness of unlearning. Smaller the metrics, better the unlearning. 3) \textit{Unlearning Time and the number of Trainable Parameters (\#Params)}: measure the efficiency of unlearning method.

\subsection{Results Analysis}

We list the comparison results in Table~\ref{tab:mainres_ml} and \ref{tab:mainres_gd}. From the results on two datasets, we observe that: 1) Our method E2URec can better maintain the recommendation performance. E2URec achieves better AUC, ACC and LogLoss compared to other baselines. This is because E2URec minimizes the KL distance between the forgetting teacher and unlearned model to remove knowledge, instead of  reversing gradients as in previous methods, thereby preserving model performance. 2) The prediction distributions of our unlearned model on forgotten data closely align with the retrained model, evidenced by the smallest JSD and L2-norm. This indicates that E2URec achieves the best unlearning effect due to our innovative forgetting teacher design, which only requires to modify the model's output to mimic the retrained model. 3) E2URec attains superior unlearning efficiency compared to other methods. E2URec has the lowest time cost and \#Params since it only updates lightweight LoRA parameters instead of all model parameters.

\subsection{Impact of LLM}
In this section, we further explore the impact of LLM on the effectiveness and efficiency. we conduct experiments on the MovieLens-1M dataset using a large LLM: T5-large (770M parameters)~\cite{t5} and a LLM with the different structure: BART-base (140M parameters)~\cite{lewis2020bart}. The results are listed in Table~\ref{tab:mainres_t5large} and~\ref{tab:mainres_bart}, from which we can find that: 1) E2URec achieves the best performance and the fastest speed on LLMs of different sizes and structures, which demonstrates the scalability of our method. 2) When the size of the LLM backbone scales up (from T5-base to T5-large), the model's recommendation performance improves significantly and the unlearning time also increases. This is because larger LLMs possess a broader range of open-world knowledge and greater reasoning abilities, but also require greater resources for finetuning.

\begin{table}[t]
    \centering
    \begin{footnotesize}
    \caption{The results (in \%) on ML-1M using T5-large. The best results (except for original and retrain) are in bold.}
    \label{tab:mainres_t5large}
    \resizebox{\linewidth}{!}{
    \renewcommand\arraystretch{1}
    \begin{tabular}{cccccccc}
    \toprule
         \multirow{2}{*}{\textbf{Metrics}} & \multicolumn{5}{c}{\textbf{Effectiveness}} & \multicolumn{2}{c}{\textbf{Efficiency}} \\ \cmidrule(r){2-6} \cmidrule(r){7-8}
         & \textbf{AUC} $\uparrow$ & \textbf{ACC} $\uparrow$ & \textbf{LL} $\downarrow$ & \textbf{JSD} $\downarrow$ & \textbf{L2-norm} $\downarrow$ & \textbf{Time(s)} $\downarrow$ & \textbf{\#Params} $\downarrow$ \\ \midrule

       Original  & 78.11 & 71.33 & 55.72 & - & - & 34080 & $7.7\times 10^8$ \\
       Retrain & 77.95 & 71.27 & 56.13 & - & - & 22005 & $7.7\times 10^8$ \\ \midrule
       SISA & 76.54 & 69.33 & 57.50 & 3.40 & 8.64 & 15210 & $3.1\times 10^9$ \\
       RecEraser & 76.78 & 69.46 & 57.77 & 1.89 & 9.74 & 15343 & $3.1\times 10^9$\\
       NegKL & 76.97 & 69.44 & 57.32 & 5.84 & 15.68 & 13458 & $7.7\times 10^8$\\
       NegGrad & 77.17 & 69.61 & 57.45 & 4.81 & 14.60 & 12960 & $7.7\times 10^8$\\
       Bad-T & 76.68 & 69.57 & 57.22 & 3.88 & 14.14 & 14049 & $7.7\times 10^8$\\
       \textbf{E2URec} & \textbf{77.52} & \textbf{70.96} & \textbf{57.02} & \textbf{1.61} & \textbf{8.25} & \textbf{8925} & \boldmath{$4.7\times 10^6$} \\
       \bottomrule
    \end{tabular}
    }
    \end{footnotesize}
\end{table}

\begin{table}[t]
    \centering
    \begin{footnotesize}
    \caption{The results (in \%) on ML-1M using BART-base. The best results (except for original and retrain) are in bold.}
    \label{tab:mainres_bart}
    \resizebox{\linewidth}{!}{
    \renewcommand\arraystretch{1}
    \begin{tabular}{cccccccc}
    \toprule
    \multirow{2}{*}{\textbf{Metrics}} & \multicolumn{5}{c}{\textbf{Effectiveness}} & \multicolumn{2}{c}{\textbf{Efficiency}} \\ \cmidrule(r){2-6} \cmidrule(r){7-8}
         & \textbf{AUC} $\uparrow$ & \textbf{ACC} $\uparrow$ & \textbf{LL} $\downarrow$ & \textbf{JSD} $\downarrow$ & \textbf{L2-norm} $\downarrow$ & \textbf{Time(s)} $\downarrow$ & \textbf{\#Params} $\downarrow$ \\ \midrule
       Original  & 76.91 & 69.97 & 57.41 & - & - & 9417 & $1.4\times 10^8$ \\
       Retrain & 76.57 & 69.92 & 57.95 & - & - & 5057 & $1.4\times 10^8$ \\ \midrule
       SISA & 74.92 & 68.71 & 59.27 & 2.92 & 11.11 & 3039 & $5.6\times 10^8$ \\
       RecEraser & 74.63 & 68.57 & 59.32 & 3.96 & 13.26 & 3273  & $5.6\times 10^8$ \\
       NegKL & 75.18 & 68.87 & 60.97 & 2.85 & 10.57 & 2583 & $1.4\times 10^8$  \\
       NegGrad & 74.93 & 67.89 & 64.18 & 5.03 & 14.40 & 2018 & $1.4\times 10^8$ \\
       Bad-T & 75.01 &68.11 & 59.93 & 4.34 & 14.24 & 2087  & $1.4\times 10^8$ \\
       \textbf{E2URec} & \textbf{76.41} & \textbf{69.95} & \textbf{57.74} & \textbf{1.43} & \textbf{7.91} & \textbf{591} & \boldmath{$8.8\times 10^5$} \\
       \bottomrule
    \end{tabular}
    }
    \end{footnotesize}
\end{table}

\subsection{Ablation Study}

We also conduct the ablation study to explore the contribution of each loss. In Table~\ref{tab:ablation}, ``w/o $L_{FGT}$'' and ``w/o $L_{REM}$'' respectively represent removing $L_{FGT}$ and $L_{REM}$. We observe that removing $L_{FGT}$ would increase JSD significantly, indicating that $L_{FGT}$ is the main factor to forget the data. Removing $L_{REM}$ would result in a notable drop in AUC, suggesting that $L_{REM}$ is essential to maintain recommendation performance. 
\begin{table}[t]
    \centering
    \begin{footnotesize}
    \caption{Ablation results (in \%). The best results are in bold.}
    \label{tab:ablation}
    \setlength{\tabcolsep}{4.4mm}{
    \begin{tabular}{ccccc}
    \toprule
    \multirow{2}{*}{\textbf{Variants}} & \multicolumn{2}{c}{\textbf{MovieLens-1M}} & \multicolumn{2}{c}{\textbf{GoodReads}} \\  \cmidrule(r){2-3}  \cmidrule(r){4-5}
         & \textbf{AUC} $\uparrow$ & \textbf{JSD} $\downarrow$ & \textbf{AUC} $\uparrow$ & \textbf{JSD} $\downarrow$ \\ \midrule
       E2URec & \textbf{76.34} & \textbf{1.91} & \textbf{73.41} & \textbf{0.90} \\
       w/o $L_{FGT}$ & 76.25 & 2.62 & 73.40 & 1.25 \\
       w/o $L_{REM}$ & 75.75 & 2.27 & 72.98 & 0.99 \\
       \bottomrule
    \end{tabular}
    }
    \end{footnotesize}
\end{table}

\section{Conclusion}
In this paper, we propose E2URec, which is the first efficient and effective unlearning method for LLMRec. Our method enables LLMRec to efficiently forget the specific data by only updating the lightweight LoRA modules. Besides, to enhance the effectiveness, our method incorporates two teacher models to guide the unlearned model in forgetting information without harming recommendation performance. Extensive experiments show that E2URec outperforms state-of-the-art baselines on two real-world datasets.

\bibliographystyle{plainnat}
\bibliography{fcs}


\end{document}